\documentclass[review]{elsarticle}
\usepackage{lineno,hyperref}
\usepackage[english]{babel}
\usepackage[numbers]{natbib}
\usepackage{xcolor}
\usepackage{latexsym,amsmath,amssymb,amsbsy,graphicx,geometry}
\modulolinenumbers[5]

\begin{document}
	
	\begin{frontmatter}
		
\title{Unified scaling model for viscosity of crude oil over extended temperature range}
		
\author[kfu]{Bulat~N.~Galimzyanov\corref{cor1}}
\cortext[cor1]{Corresponding author}
\ead{bulatgnmail@gmail.com}

\author[kfu]{Maria~A.~Doronina}
\ead{maria.doronina.0211@gmail.com}
	
\author[kfu]{Anatolii~V.~Mokshin}
\ead{anatolii.mokshin@mail.ru}
		
\address[kfu]{Kazan Federal University, 420008 Kazan, Russia}

\begin{abstract}
The viscosity of crude oil is an important physical property that largely determines the fluidity of oil and its ability to seep through porous media such as geological rock. Predicting crude oil viscosity requires the development of reliable models that can reproduce viscosity over a wide range of temperatures and pressures. Such viscosity models must operate with a set of physical characteristics that are sufficient to describe the viscosity of an extremely complex multi-phase and multi-component system such as crude oil. The present work considers empirical data on the temperature dependence of the viscosity of crude oil samples from various fields in Russia, China, Saudi Arabia, Nigeria, Kuwait and the North Sea. For the first time, within the reduced temperature concept and using the universal scaling viscosity model, the viscosity of crude oil can be accurately determined over a wide temperature range: from low temperatures corresponding to the amorphous state to relatively high temperatures, at which all oil fractions appear as melts. A novel methodology for determining the glass transition temperature and the activation energy of viscous flow of crude oil is proposed. A relationship between the parameters of the universal scaling model for viscosity, the API gravity, the fragility index, the glass transition temperature and the activation energy of viscous has been established for the first time. It is shown that the accuracy of the results of the universal scaling model significantly exceeds the accuracy of known empirical equations, including those developed directly to describe the viscosity of petroleum products.
\end{abstract}


\end{frontmatter}

\section{Introduction}

\begin{figure}[ht!]
	\centering
	\includegraphics[width=1.0\linewidth]{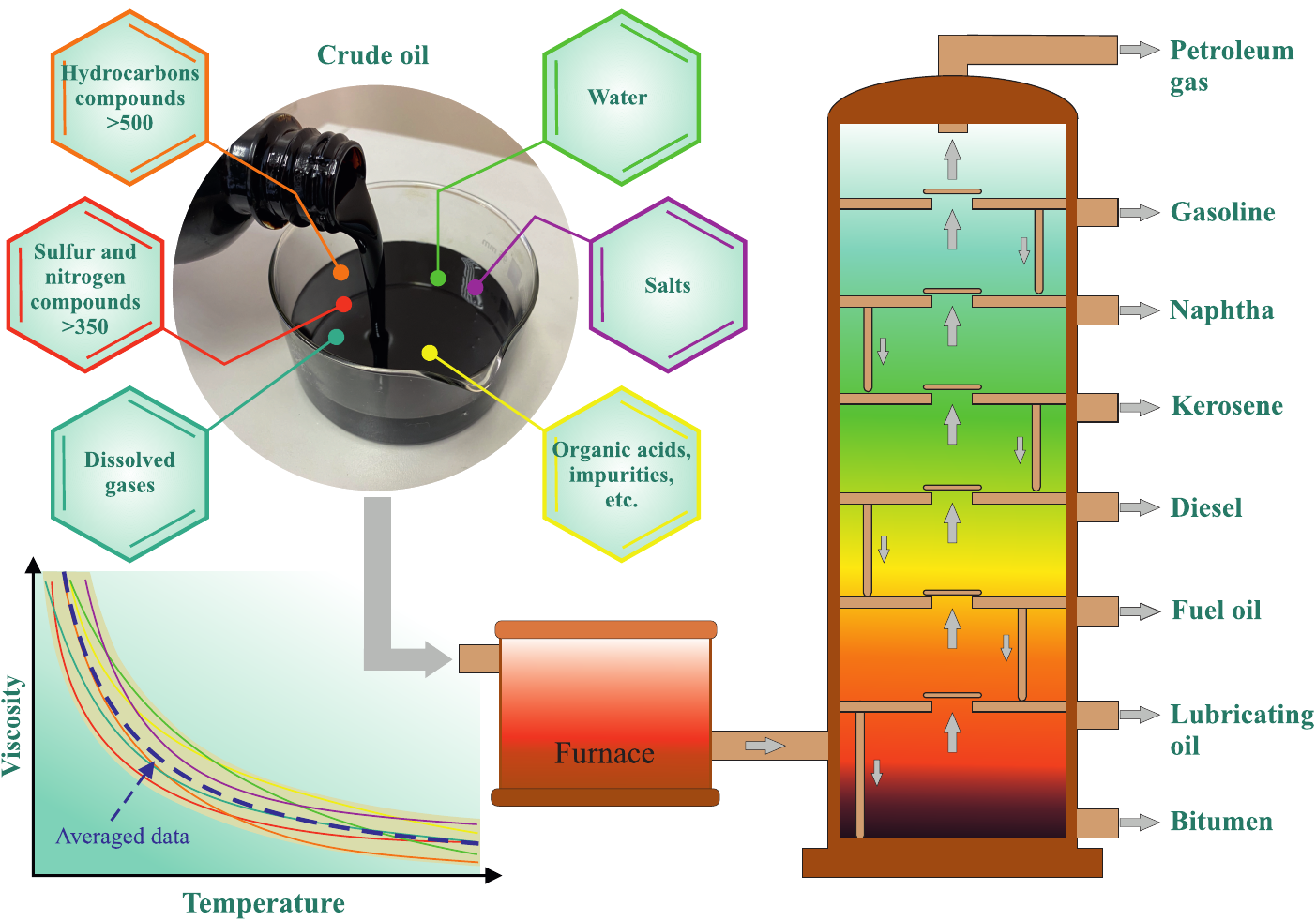}
	\caption{Basic composition of crude oil and scheme of the process for extracting the fractions. The viscosity of crude oil is determined by the viscosity of the individual compounds.}
	\label{fig_1}
\end{figure}

Crude oil is one of the most valuable resource that is used in the world. It consists of a complex mixture of liquid hydrocarbons of different molecular weights (more than $500$ compounds; $80$--$90$\%), such as paraffins, naphthenes, aromatic hydrocarbons, and also contains sulphur, nitrogen and oxygen compounds (more than $350$ compounds; up to $5$\%), water (up to $10$\%), dissolved gases (up to $4$\%), mineral salts, organic acids, chelate complexes and impurities [see Figure~\ref{fig_1}]. In total, crude oil contains about $1000$ different compounds. Each compound has its own set of physical and chemical properties~\cite{Douglas_2022}. Therefore, crude oil can be regarded as the prototype of a ``complex liquid'' with a extremely complex heterogeneous structure and complex composition.

The physical and chemical properties of crude oil are unique to an each oil field. Some of the most significant physical properties of crude oil are the fractional composition, the viscosity, the density, the API gravity, the average molecular weight and others. For various technological applications related to enhanced oil recovery, oil production and transportation, it is particularly important to predict the composition-related physical properties of crude oil. Such properties include the viscosity, which determines the fluidity of oil under certain thermodynamic conditions. The presence of heavy fractions such as paraffins, ceresins, asphaltenes and resins makes oil highly viscous and more difficult to extract~\cite{Shah_Fishwick_2010}. Increasing the proportion of light fractions such as petroleum, gasoline, diesel, etc., reduces the viscosity of oil and thus makes it easier to extract~\cite{Luo_Gu_2007,Guliaeva_Smirnov_2020}. Correct estimation of the crude oil viscosity is one of the important tasks of the petroleum industry, especially in the field of extraction of hard-to-recover high-viscosity reserves.

Accurate determination of the crude oil viscosity is of great importance for correct simulation and prediction of oil flow in various media such as porous oil reservoirs, boreholes and pipelines. Due to the diverse composition and geographical characteristics of oil reservoirs, accurate viscosity models are required to predict the temperature dependence of the crude oil viscosity~\cite{Alomair_Jumaa_2016}. Microscopic viscosity models such as the nonaffine theory of viscoelasticity~\cite{Zaccone_2023}, which describe the temperature dependence of the viscosity in terms of the local environment of atoms/molecules, the molecular cooperation and the molecular density of vibrational states, are not applicable in the case of crude oil due to its strong structural heterogeneity. Viscosity models that describe the viscosity in terms of macroscopic physical parameters, such as the empirical models with the Vogel-Fulcher-Tamman equation~\cite{Nascimento_Aparicio_2007}, the Masuko-Magill~\cite{Masuko_Magill_1988} and William-Landel-Ferry~\cite{Kolotova_Kuchina_2018, Williams_Landel_Ferry_1955} models, are usually effective only in the case of pure single-type compounds belonging to the class ``fragile'' according to the Angell's classification for glass-formers~\cite{Kelton_2023}. In the case of mixtures and multiphase systems, these models do not correctly describe the temperature dependence of the viscosity over an extended temperature range. There are also viscosity models that have been developed directly for oil: for example, the Alomair et al. model, the Orbey and Sandler model~\cite{Alomair_Jumaa_2016,Bergman_2007}. The equations of these models include the API gravity, the average molecular weight and/or the density of oil~\cite{Sanchez-Minero_2014}. These models are usually applicable to a very limited range of temperatures, and, in addition, give incorrect results in the case of oil with highly pronounced structural and dynamic heterogeneities. Machine learning viscosity models have also been proposed~\cite{Rodriguez_Galiano_2015,Pugliese_Regondi_2021,Li_Zhang_2023}. The training and correct use of the machine learning viscosity models requires a large set of high quality data that takes into account the physical and chemical properties of each oil fraction. Therefore, the generation of such datasets requires a careful analysis of the oil composition and accurate systematization of the data. Therefore, empirical and analytical models of crude oil viscosity are still required.

One of the ways to solve the problem of developing a viscosity model capable of correctly describing the viscosity of crude oil, among others, is to use scaling concepts. Scaling concepts aim to describe the complex behavior of a system by representing its physical properties as a function of a single variable, the value of which can be accurately estimated experimentally and/or from simulation results. Concepts such as the excess entropy scaling~\cite{Bell_Dyre_2020} and the density scaling~\cite{Pawlus_Grzybowski_2020} have demonstrated their applicability in the development of models for shear viscosity, self-diffusion coefficient and thermal conductivity of pure liquids, mixtures and active matter~\cite{Fragiadakis_Roland_2019,Ghaffarizadeh_Wang_2022}. These concepts use quantities directly related to interatomic interactions and structure. It is limits their application to crude oil, which is characterized by complex interparticle interactions and extremely complex structure. Previously, we have developed the concept of temperature scaling and demonstrated its applicability to construct the unified scaling viscosity model capable of reproducing the viscosity of different systems over a wide temperature range~\cite{Galimzyanov_Mokshin_UVM_2021,Mokshin_Galimzyanov_JCP_2015}. This model has shown high accuracy compared to other analytical and empirical viscosity models in reproducing empirical data on the viscosity of silicate, borate, metal and organic melts. In the present work, this model is developed to describe the viscosity of crude oil.

The main aim of the present study is to describe the temperature dependence of the viscosity and to evaluate the glass transition  characteristics of crude oil. The proposed unified scaling model for viscosity is tested on a set of empirical data for oil from various fields in Russia, China, Saudi Arabia, Kuwait, Nigeria and the North Sea. The ability of this model to reproduce the viscosity of crude oil over a wide temperature range, including low temperatures at which oil becomes highly viscous due to amorphisation of solid fractions, is demonstrated.

\section{Methods and Materials}

\begin{figure}[ht!]
	\centering
	\includegraphics[width=1.0\linewidth]{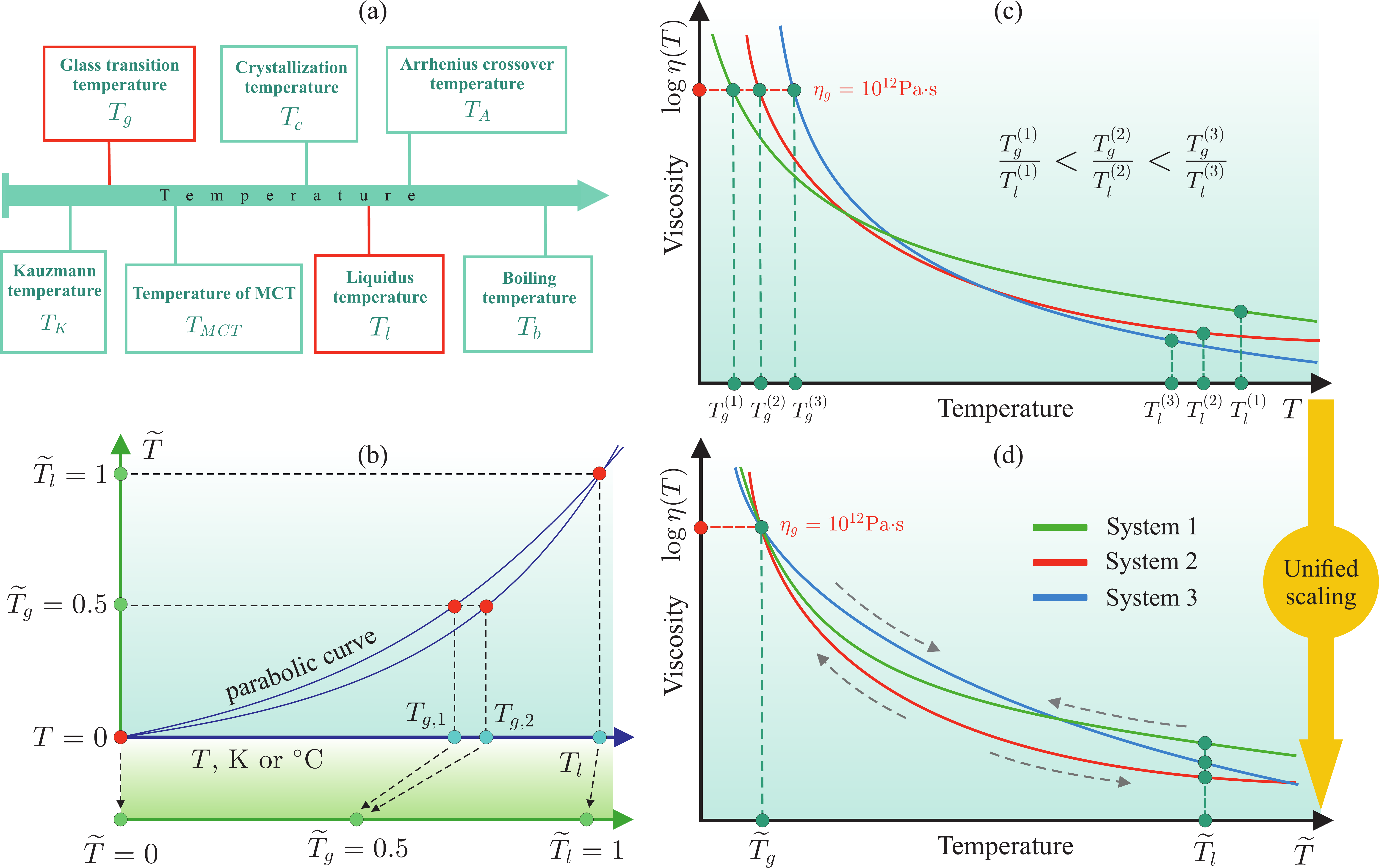}
	\caption{(a) Special temperatures characterizing the state of a physical system. (b) Schematic representation of the transition from the absolute temperature scale $T$ to the reduced temperature scale $\widetilde{T}$. (c) Example of the temperature dependence of the viscosity of three arbitrary systems with different ratio $T_{g}/T_{l}$ in the absolute temperature scale $T$. (d) Example of the temperature dependence of the viscosity of three arbitrary systems, where the reduced temperature $\widetilde{T}$ is used.}
	\label{fig_2}
\end{figure}

Any liquid is characterized by a set of so-called ``special temperatures'' related to changes in the thermodynamics of a liquid [see Figure~\ref{fig_2}(a)]. Among those temperatures of special interest are the glass transition temperature $T_{g}$ and the liquidus temperature $T_{l}$, which are directly related to the change in the fluidity of a liquid. The width of the temperature range [$T_{g}$;\,$T_{l}$] (or the value of the temperature ratio $T_{g}/T_{l}$) depends on the composition of a liquid. For example, for propylene glycol (C$_3$H$_8$O$_2$) one has $T_{g}/T_{l}\approx0. 77$, for triacontane (C$_{30}$H$_{62}$) one finds $T_{g}/T_{l}$ is $\approx0.72$ and for pentanal (C$_5$H$_{10}$O) one obtains $T_{g}/T_{l}\approx0.67$~\cite{Jaiswal_Egami_2016}.

From the above, it is reasonable to assume that in order to correctly describe the temperature dependence of physical properties (including the viscosity $\eta$) of different type systems, it is necessary to use a universal temperature scale, which represents the change in physical properties in a uniform way. A viscosity model that claims to be universal should correctly reproduce the temperature dependence of the viscosity of a system in a wide temperature range: starting from temperatures, at which the viscosity of oil is extremely high and oil itself is a multiphase system, where some fractions are represented in solid state and some -- in a liquid state. For this purpose, a viscosity model must takes into account the thermodynamic state of a system based on ``special temperatures'' such as the glass transition temperature $T_{g}$ and the liquidus temperature $T_{l}$. Previously, we have proposed the unified scaling model for viscosity (USMV), which is able to uniformly reproduce the temperature dependence of the viscosity of different type systems~\cite{Galimzyanov_Mokshin_UVM_2021}:
\begin{equation}\label{eq_viscosity_model}
\log\eta(\widetilde{T})=\log\eta_{\infty}+\alpha\left(\frac{\widetilde{T}_{g}}{\widetilde{T}}\right)^{\beta},\quad \alpha=\log\left(\frac{\eta(T_{g})}{\eta_{\infty}}\right).
\end{equation}
Here, $\eta_{\infty}$ is the high temperature limit ($T\rightarrow\infty$) of the viscosity, $\eta(T_{g})$ is the viscosity at the glass transition temperature $T_{g}$, where we have $\eta(T_{g})=10^{12}$\,Pa$\cdot$s. The exponent $\beta$ is positive and for most glass-forming liquids takes values in the range $\beta\in[0.5;\,5.0]$~\cite{Galimzyanov_Mokshin_UVM_2021}. In the case of liquids belonging to the class of ``strong'' glass-formers according to the Angell's classification~\cite{Angell_1995}, we have $0.5<\beta\leq1.0$. In the case of systems belonging to the ``fragile'', the exponent is $\beta>1.0$~\cite{Galimzyanov_Mokshin_UVM_2021}. Here, the value $\beta=0.5$ corresponds to the limiting case when the temperature dependence of the viscosity of a system is described over the wide temperature range up to the glass transition temperature by the Frenkel-Andrade equation~\cite{Shirai_2021,Frenkel_1946}:
\begin{equation}\label{eq_Arr1}
\eta(T)=\eta_{\infty}\exp\left(\frac{E_{A}}{RT_{g}}\right),
\end{equation}
where $R\simeq8.31$\,J/(mol$\cdot$K) is the universal gas constant. In Eq.~(\ref{eq_Arr1}), the activation energy $E_{A}$ is the temperature dependent quantity near the glass transition temperature, while for most liquids we have $E_{A}$=const at high temperatures away from $T_{g}$~\cite{Sagdeev_Isyanov_2020}. Then, it can be shown from Eqs.~(\ref{eq_viscosity_model}) and (\ref{eq_Arr1}) that the value of the quantity $\alpha$ is related to the activation energy $E_{A}^{(T_g)}$ of viscous flow at the glass transition temperature $T_{g}$~\cite{Galimzyanov_Mokshin_UVM_2021}:
\begin{equation}\label{eq_alpha_E}
\alpha=\frac{1}{\ln 10}\frac{E_{A}^{(T_g)}}{RT_{g}}.
\end{equation}  

In Eq.~(\ref{eq_viscosity_model}), the reduced temperature $\widetilde{T}$ is used instead of the absolute temperature, which is defined through the following expression~\cite{Galimzyanov_Mokshin_UVM_2021,Mokshin_Galimzyanov_JCP_2015,Mokshin_Galimzyanov_JETPL_2019}:
\begin{equation}\label{eq_scaled_tmp}
\widetilde{T}=K_{1}(T_{g},T_{l})\left[\frac{T}{T_{g}}\right]+K_{2}(T_{g},T_{l})\left[\frac{T}{T_{g}}\right]^{2},
\end{equation}
where
\begin{equation}
K_{1}(T_{g},T_{l})=\left(0.5-\left[\frac{T_g}{T_l}\right]^{2}\right)\left(1-\frac{T_g}{T_l}\right)^{-1}, \nonumber
\end{equation}
\begin{equation}
K_{2}(T_{g},T_{l})=\frac{T_g}{T_l}\left(\frac{T_g}{T_l}-0.5\right)\left(1-\frac{T_g}{T_l}\right)^{-1}.\nonumber
\end{equation}
Here, $K_{1}(T_{g},T_{l})$ and $K_{2}(T_{g},T_{l})$ are the coefficients dependent on $T_{g}$ and $T_{l}$. The sum of these coefficients is $K_{1}(T_{g},T_{l})+K_{2}(T_{g},T_{l})=0.5$. The expressions for the coefficients $K_{1}(T_{g},T_{l})$ and $K_{2}(T_{g},T_{l})$ are obtained strictly based on the requirement that the calibration of the temperature scale is fulfilled and that the values of the glass transition temperature $T_{g}$ and the liquidus temperature $T_{l}$ are equal to $0.5$ and $1.0$, respectively. This temperature scaling is performed for the temperature range from ultra-low temperatures comparable with $T=0$ to temperatures corresponding to the equilibrium melt. 

Figure~\ref{fig_2}(b) shows that an arbitrary value of the liquidus temperature $T_{l}$ in the Kelvin scale corresponds to the fixed value $\widetilde{T}_{l}=1.0$ in the reduced temperature scale, while the glass transition temperature $T_{g}$ in the reduced temperature scale is always $\widetilde{T}_{g}=0.5$. The liquidus and glass transition lines on the ($p$, $\widetilde{T}$) phase diagram will be represented as parallel isotherms $\widetilde{T}_{l}=1.0$ and $\widetilde{T}_{g}=0.5$ (here, $p$ is the pressure). Thus, the use of the reduced temperature $\widetilde{T}$ makes it possible to represent the temperature dependences of the viscosity of different type systems in a unified way [see Figures~\ref{fig_2}(c) and~\ref{fig_2}(d)]. 

Using (\ref{eq_alpha_E}) and (\ref{eq_scaled_tmp}), Eq.~(\ref{eq_viscosity_model}) for the USMV can be represented as follows:
\begin{equation}\label{eq_viscosity_model_Tk}
\log\eta(T)=\log\eta(T_g)+\frac{1}{\ln 10}\frac{E_{A}^{(T_g)}}{RT_{g}}\left\{\left(2K_{1}(T_{g},T_{l})\left[\frac{T}{T_{g}}\right]+2K_{2}(T_{g},T_{l})\left[\frac{T}{T_{g}}\right]^{2}\right)^{-\beta}-1\right\}.
\end{equation}
Eq.~(\ref{eq_viscosity_model_Tk}) reproduces the viscosity of a system as a function of temperature $T$. A quantitative characteristic of the viscosity change with temperature in the vicinity of the temperature $T_{g}$ is the so-called fragility index
\begin{equation}\label{eq_fragindex}
m=\frac{\partial\log\eta(T)}{\partial(T_g/T)}\Bigg|_{T=T_{g}},
\end{equation}
which indicates how rapidly the atomic dynamics slow down near the glass transition temperature $T_{g}$~\cite{Angell_1995,Qin_McKenna_2006,Zheng_Mauro_2017}. The lower the value of the fragility index $m$, the slower the viscosity changes with temperature in the vicinity $T_g$. From Eqs.~(\ref{eq_viscosity_model_Tk}) and (\ref{eq_fragindex}), we find
\begin{equation}\label{eq_model_fragindex}
m=\frac{\beta}{\ln 10}\frac{E_{A}^{(T_g)}}{RT_{g}}[1+2K_{2}(T_{g},T_{l})].
\end{equation}
This expression makes it possible to determine the fragility index from the known values of the activation barrier $E_{A}^{(T_g)}$ and the exponent $\beta$. Typically, the fragility index takes values in the range from $m\simeq17$ (for example, SiO$_{2}$, GeO$_{2}$) to $m\simeq120$ (for example, polymers, complex hydrocarbons)~\cite{Debenedetti_Stillinger_2001,Avramov_2005}. In the present work, the USMV represented as (\ref{eq_viscosity_model_Tk}) will be used to describe the viscosity-temperature data obtained for the crude oil samples.

This work considers twenty crude oil samples from different countries and fields, including two samples from Russia (Republic of Tatarstan), four samples from China (Tarim Basin and Junggar Basin), six samples from Nigeria (Ondo State and Niger Delta area), six samples from Kuwait and one sample each from Saudi Arabia and the North Sea region (see Table~\ref{tab_1}). These samples have a unique combinations of physical properties including the density $\rho$~\footnote{The density of crude oil depends on its content of heavy hydrocarbons (paraffins, resins, etc.). Crude oil is usually classified as light at $\rho<870$\,kg/m$^{3}$, medium at densities from $870$\,kg/m$^{3}$ to $970$\,kg/m$^{3}$ and heavy at $\rho>970$\,kg/m$^{3}$.}, the glass transition, the liquidus and boiling temperatures at normal pressure, the API gravity~\footnote{The API gravity is the unit of measurement of petroleum density developed by the American Petroleum Institute. This parameter determines the oil density relative to the water density at the temperature $289$\,K. Typically, the API gravity takes values between $3$ and $60$: $\text{API}>31$ for light oil, $\text{API}\in[22;\,31]$ for medium oil, $\text{API}<22$ for heavy oil. The API gravity of crude oil is determined by the Standard Test Method for API Gravity of Crude Petroleum and Petroleum Products (ASTM)~\cite{ASTM_2024}.} and the average molecular weight~\footnote{This defines the total molecular weight of the oil individual components and takes values in the range $M\in[220;\,300]$\,g/mol. Usually, the Voinov's formula is used for the simplified determination of $M$: $M=60+0.3\bar{T}_{b}+0.001\bar{T}_{b}^{2}$, where $\bar{T}_{b}$ is the average boiling temperature of oil.}.

\section{Results and Discussion}

\begin{figure*}[ht!]
	\centering
	\includegraphics[width=0.8\linewidth]{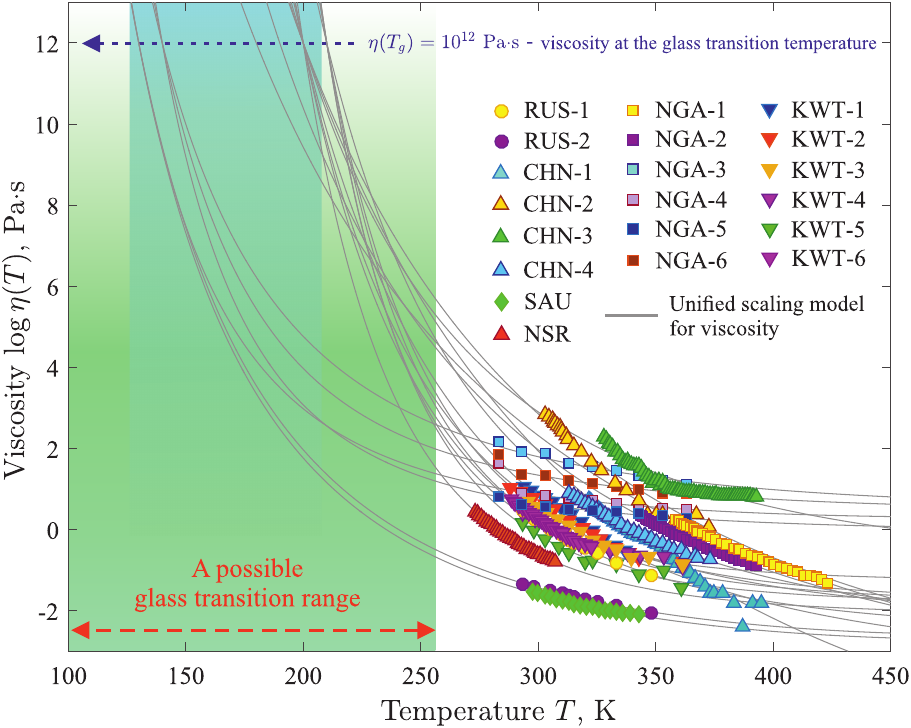}
	\caption{Temperature dependence of the crude oil viscosity $\eta(T)$ for the considered samples. The coloured markers are the experimental data. The solid lines are the results of the unified scaling model for viscosity obtained by Eq.~(\ref{eq_viscosity_model_Tk}). A possible region for the glass transition temperature is indicated by the red arrow.}
	\label{fig_3}
\end{figure*}

Figure~\ref{fig_3} shows the temperature dependence of the viscosity $\eta(T)$ for the considered oil samples. The viscosity-temperature data for each sample are very different, mainly due to differences in the proportion of solid fractions in the oil composition. The viscosity measurements were made near the temperatures, at which the dense oil fractions such as paraffin and ceresin are melted. Thus, these fractions transit to the liquid phase at temperatures $325\pm25$\,K~\cite{Sathivel_King_2008}. Therefore, in the present work we define the liquidus temperature of crude oil as $T_{l}\simeq350$\,K, which corresponds to the temperature at which the paraffin fractions melt completely~\cite{Bai_Fan_2019}.

According to the generally accepted definition, the glass transition temperature $T_{g}$ is the temperature at which the viscosity of a system is $\eta=10^{12}$\,Pa$\cdot$s, and the structural relaxation time takes the value $100$\,s~\cite{Sanditov_Darmaev_2017,Buchholz_Binder_2002}. According to the available empirical data and the results obtained using machine learning models, the glass transition temperature of crude oil can range from $T_g\sim100$\,K to $T_g\sim260$\,K ~\cite{Masson_Polomark_2006,Claudy_1988,Kutcherov_1993,Galimzyanov_Doronina_2023}. In order to determine the glass transition temperature $T_{g}$, the crude oil viscosity was reproduced using the USMV represented by Eq.~(\ref{eq_viscosity_model_Tk}). In this equation, the temperature $T_{g}$, the activation energy $E_{A}^{(T_g)}$ and the exponent $\beta$ were taken as fitting parameters. The result of reproducing the fitting procedure is shown in Figure~\ref{fig_3} (solid curves). The found values of the glass transition temperature $T_{g}$ are in the range $T_g\in[100;\,260]$\,K (see Table~\ref{tab_1})~\cite{Claudy_1988,Kutcherov_1993}. Crude oil samples extracted from Nigerian fields have a wide spread of the glass transition temperatures, $T_{g}\in[130;\,180]$\,K. This may be due to the significant differences in the structure and composition of these samples. This is evidenced by the fact that the density of these samples varies from $0.843$\,g/cm$^3$ to $0.933$\,g/cm$^3$~\cite{Akankpo_Essien_2015}. Such differences are also observed in the glass transition temperatures of crude oil from fields in the Republic of Tatarstan (Russia): for the Ashal'chinskaya oil field (marked as RUS-1) with the glass transition temperature $T_{g}\simeq200$\,K and for the Kuakbashskaya oil field (marked as RUS-2) with $T_{g}\simeq130$\,K. Sample RUS-1 has a high content of heavy oil fractions such as resin (28.8~wt\,\%) and asphaltene (6.52~wt\,\%) as well as water (0.4~wt\,\%). Sample RUS-2 has a lower content of resin (16.3~wt\,\%), asphaltene (4.12~wt\,\%) and water (traces)~\cite{Sagdeev_Isyanov_2020}. Oil samples from China, Saudi Arabia, Kuwait and the North Sea region have the glass transition temperature $T_{g}\simeq205\pm5$\,K, which may indicate a similar composition.

\begin{table}[tbh]
	\scriptsize
	\centering
	\caption{Characteristics of the considered crude oil samples: brief designation and name of the oil production site (basin); the API gravity; the glass transition temperature $T_{g}$; the activation energy $E_{A}^{(T_g)}$ at the temperature $T_{g}$; the parameters $\beta$ and $K_{2}(T_g,\,T_l)$ of Eq.~(\ref{eq_viscosity_model_Tk}); the fragility index $m$. The last column indicates the literature sources, where the experimental viscosity data were taken. For the quantities $T_{g}$ and $E_{A}^{(T_g)}$ the estimation error is $\sim8$\,\%.}	
	\begin{tabular}{ccccccccc}
		\hline
		Name & Description & API & $T_{g}$, K & $E_{A}^{(T_g)}$, kJ/mol & $\beta$ & $K_{2}(T_g,\,T_l)$ & $m$ & Refs. \\
		\hline
		RUS-1 & Russia, Tatarstan 			& $16.2\pm6.8$ & $200$ & $55.9$ & $3.32\pm0.4$ & $0.09524$ & $57.7\pm5$ & \cite{Sagdeev_Isyanov_2020} \\
		RUS-2 & Russia, Tatarstan 			& $25.2\pm4.8$ & $130$ & $41.5$ & $2.66\pm0.3$ & $-0.076$ & $37.6\pm3$ & \cite{Sagdeev_Isyanov_2020} \\
		CHN-1 &	China, Tarim Basin 			& $17.5\pm4.5$ & $200$ & $78.3$ & $1.37\pm0.2$ & $0.09524$ & $33.2\pm3$ & \cite{Chen_Pu_2018} \\
		CHN-2 & China, Tarim Basin			& $10.0\pm5.0$ & $200$ & $67.0$ & $1.49\pm0.2$ & $0.09524$ & $31.0\pm3$ & \cite{Qin_Wu_2018} \\
		CHN-3 &	China, Tarim Basin 			& $4.43\pm2.6$ & $200$ & $50.2$ & $2.39\pm0.2$ & $0.09524$ & $37.3\pm4$ & \cite{Jin_Jiang_2022} \\
		CHN-4 &	China, Junggar Basin 		& $19.2\pm2.8$ & $210$ & $58.9$ & $2.62\pm0.3$ & $0.15$ & $49.9\pm8$ & \cite{Jia_Liu_2016} \\
		SAU	  & Saudi Arabian 				& $27.0\pm5.0$ & $130$ & $41.3$ & $2.77\pm0.4$ & $-0.076$ & $39.0\pm4$ & \cite{Karnanda_Benzagouta_2013} \\
		NSR	  & North Sea Region 			& $19.0\pm5.0$ & $200$ & $54.1$ & $4.50\pm0.4$ & $0.09524$ & $75.7\pm6$ & \cite{Kolotova_Kuchina_2018} \\
		NGA-1 &	Nigeria, Ondo State 		& $7.20\pm1.8$ & $180$ & $66.1$ & $1.32\pm0.2$ & $0.01513$ & $26.2\pm3$ & \cite{Alade_Shehri_2019} \\
		NGA-2 &	Nigeria, Ondo State 		& $8.60\pm1.8$ & $190$ & $60.7$ & $1.79\pm0.2$ & $0.5089$ & $32.9\pm4$ & \cite{Alade_Shehri_2019} \\
		NGA-3 &	Nigeria, Niger Delta area 	& $20.1\pm3.9$ & $140$ & $32.8$ & $3.01\pm0.6$ & $-0.0667$ & $31.9\pm5$ & \cite{Akankpo_Essien_2015} \\
		NGA-4 & Nigeria, Niger Delta area	& $25.1\pm6.2$ & $130$ & $30.7$ & $3.81\pm0.6$ & $-0.076$ & $39.3\pm4$ & \cite{Akankpo_Essien_2015} \\
		NGA-5 & Nigeria, Niger Delta area	& $36.4\pm6.6$ & $140$ & $32.8$ & $4.36\pm0.5$ & $-0.0667$ & $46.4\pm6$& \cite{Akankpo_Essien_2015} \\
		NGA-6 & Nigeria, Niger Delta area	& $32.5\pm5.5$ & $140$ & $31.9$ & $3.75\pm0.5$ & $-0.0667$ & $38.8\pm4$ & \cite{Akankpo_Essien_2015} \\
		KWT-1 &	Kuwait 						& $18.8\pm3.2$ & $200$ & $59.4$ & $2.57\pm0.2$ & $0.09524$ & $47.5\pm4$ & \cite{Alomair_Jumaa_2016} \\
		KWT-2 &	Kuwait 						& $18.8\pm3.4$ & $200$ & $54.8$ & $3.19\pm0.2$ & $0.09524$ & $54.3\pm5$ & \cite{Alomair_Jumaa_2016} \\
		KWT-3 &	Kuwait 						& $15.3\pm3.7$ & $210$ & $55.6$ & $3.37\pm0.3$ & $0.15$ & $66.8\pm5$ & \cite{Alomair_Jumaa_2016} \\
		KWT-4 &	Kuwait 						& $15.3\pm3.7$ & $210$ & $53.4$ & $4.64\pm0.4$ & $0.15$ & $80.2\pm7$ & \cite{Alomair_Jumaa_2016} \\
		KWT-5 &	Kuwait 						& $11.7\pm3.3$ & $210$ & $55.7$ & $4.39\pm0.4$ & $0.15$ & $78.9\pm6$ & \cite{Alomair_Jumaa_2016} \\
		KWT-6 &	Kuwait 						& $14.8\pm3.2$ & $210$ & $55.6$ & $3.96\pm0.2$ & $0.15$ & $71.2\pm7$ & \cite{Souas_Safri_2021} \\
		\hline
	\end{tabular}\label{tab_1}
\end{table}

Now we are aimed to determine the correspondence between the exponent $\beta$ in Eq.~(\ref{eq_viscosity_model_Tk}) and some known crude oil characteristics. The exponent $\beta$ depends on how the viscosity of crude oil changes over a wide temperature range, including the high temperature region (near the liquidus temperature and above) and the low temperature region (in the vicinity of the glass transition temperature). This quantity can be related to the properties that implicitly characterize the composition of a system. One such parameter is the API gravity. The API gravity values are taken from~\cite{Alomair_Jumaa_2016,Kolotova_Kuchina_2018,Sagdeev_Isyanov_2020,Chen_Pu_2018,Qin_Wu_2018,Jin_Jiang_2022,Jia_Liu_2016,Karnanda_Benzagouta_2013,Souas_Safri_2021} and are given in Table~\ref{tab_1}. For the oil samples, the fragility index $m$ was estimated by Eq.~(\ref{eq_model_fragindex}) using the evaluated values of the parameters $\beta$, $E_{A}^{(T_g)}/T_{g}$ and $K_{2}(T_g,\,T_l)$ (see Table~\ref{tab_1}). Thus, for the oil samples, we have the following set of the characteristics:
\begin{equation}
\mathcal{H}=\{\text{API}, m, T_{g}, E_{A}^{(T_g)}, \beta, K_{2}(T_g,\,T_l)\}.\nonumber
\end{equation}

\begin{figure}[ht!]
	\centering
	\includegraphics[width=1.0\linewidth]{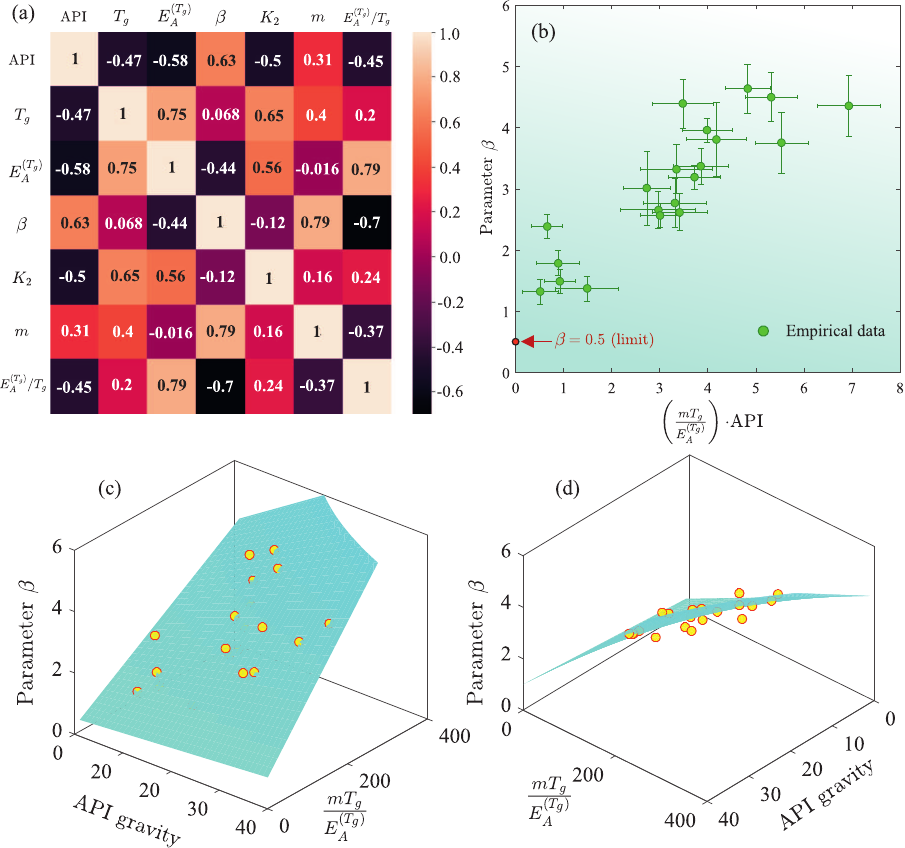}
	\caption{(a) Pearson correlation heat map for the physical characteristics of the crude oil samples. (b) Correspondence between the exponent $\beta$ and the quantity $\text{API}\cdot(mT_{g}/E_{A}^{(T_g)})$. (c) Correspondence between the parameters $\beta$, API gravity and $mT_{g}/E_{A}^{(T_g)}$. The circle shows empirical data. These data are compared with the result of Eq.~(\ref{eq_nrm_1}), which is shown as a curved surface. (d) This 3D-plot is from an another foreshortening.}
	\label{fig_4}
\end{figure}

To determine the relationship between any two characteristics in the set $\mathcal{H}$, the linear Pearson correlation coefficient was determined\footnote{The Pearson correlation coefficient is the dimensionless index that takes values between $-1.0$ and $1.0$ inclusive~\cite{Jetly_Chaudhury_2021}. Here, the value $-1.0$ indicates a fully linear inverse relationship; the value $1.0$ -- a fully linear direct relationship; $0$ -- no linear correlation.}. Figure~\ref{fig_4}(a) shows that the Pearson correlation coefficient higher than $0.6$ between $\beta$, API gravity and $m$ as well as lower than $-0.7$ between $\beta$ and $E_{A}^{(T_g)}/T_{g}$. A weak correlation is observed for other characteristics, where the Pearson correlation coefficient close to zero. Thus, on the basis of the obtained results, we have the following relationship
\begin{equation}\label{eq_propto_p}
\beta\propto\text{API}\cdot\frac{mT_{g}}{E_{A}^{(T_g)}}.
\end{equation}

Figure~\ref{fig_4}(b) shows the correspondence between the exponent $\beta$ and the combination of parameters $\text{API}\cdot(mT_{g}/E_{A}^{(T_g)})$. The scatter in the data observed in this figure indicates on a complex relationship between these parameters. From the regression analysis of the empirical data we find
\begin{equation}\label{eq_nrm_1}
\beta(K,\text{API})=a_{0}+\left(a_{1}+a_{3}K\right)\cdot\text{API}+a_{2}K,\quad K=\frac{mT_{g}}{E_{A}^{(T_g)}}.
\end{equation}
Here, $\beta$ is a function of the quantity $mT_{g}/E_{A}^{(T_g)}$ and the API gravity~\cite{Mokshin_Sharnin_2019,Mokshin_Mirziyarova_2020}. Details of the derivation of Eq.~(\ref{eq_nrm_1}) are given in the Supplementary Material. The fitting coefficients take the values $a_{0}=0.5$, $a_{1}=0.01241$, $a_{2}=0.0089$\,kJ/(K$\cdot$mol) and $a_{3}=0.00023$\,kJ/(K$\cdot$mol), respectively. Then, $a_{0}=0.5$ is the limit value for the exponent $\beta$~\cite{Galimzyanov_Mokshin_UVM_2021}. As it has been established, all other fitting coefficients are zero. With such the values of the fitting coefficients we obtain the minimum error between the calculated $\beta$ and the result of Eq.~(\ref{eq_nrm_1}). Note that mathematically this equation represents a second order curvilinear surface equation. Figures~\ref{fig_4}(c) and~\ref{fig_4}(d) show that Eq.~(\ref{eq_nrm_1}) correctly determines the correspondence between the calculated values of the parameters $\beta$, API gravity and $mT_{g}/E_{A}^{(T_g)}$. Note that the deviation from this surface is due to the presence of extra heavy (low API) and extra light (high API) oil samples. The obtained Eq.~(\ref{eq_nrm_1}) can be used to determine the physical properties of crude oil from different fields based on the results of Eq.~(\ref{eq_viscosity_model_Tk}) for the USMV.

Assume that the oil viscosity $\eta$ is related to the self-diffusion coefficient $D$ by the generalized Stokes-Einstein relation~\cite{Wei_Evenson_2018}:
\begin{equation}\label{eq_ES}
D=\left(\frac{CT}{\eta}\right)^{\xi},\,0<\xi\leq1,
\end{equation}
where $C$ is the positive constant with dimensionality [Pa$\cdot$m$^2$/K]. Recently, using the surface self-diffusion coefficient obtained for the case of crystallisation of molecular glasses, we have received the following relation~\cite{Mokshin_Galimzyanov_JETPL_2019}:
\begin{equation}\label{eq_ESmKsi}
m=\frac{1}{\ln 10}\left\{\frac{\chi}{\xi}\left[1+2K_{2}(T_{g},\,T_{l})\right]-1\right\},
\end{equation}
which relates the fragility index $m$ to the parameters $\xi$ and  $K_{2}(T_{g},\,T_{l})$. Here, parameter $K_{2}(T_{g},\,T_{l})$ can be considered as a criterion for the glass-forming ability of a system. The parameter $\chi$ characterizes the change of the self-diffusion within the temperature range ($0$, $T_l$]. In the case of the self-diffusion in the bulk of a system, one has $\chi\simeq m-17$, while in the case of the surface self-diffusion, one obtains $\chi\simeq(m-17)/3$~\cite{Mokshin_Galimzyanov_JETPL_2019}. For the oil samples, the calculations are performed for the bulk of a system. Then, taking into account that $\chi\simeq m-17$ and  relation (\ref{eq_ESmKsi}), it follows that the exponent in the Stokes-Einstein relation takes values in the range $0.17<\xi<0.45$. The breakdown of the Stokes-Einstein relation, at which $\xi\neq1$, is due to oil is characterized by structural and dynamical heterogeneities. Previously, it was shown that the Stokes-Einstein relation can be violated in the case of ``fragile'' systems, where $\xi$ can take values less than $0.6$~\cite{Ediger_Harrowell_2008}. This conclusion is consistent with the results of the present study.

\begin{figure}[ht!]
	\centering
	\includegraphics[width=1.0\linewidth]{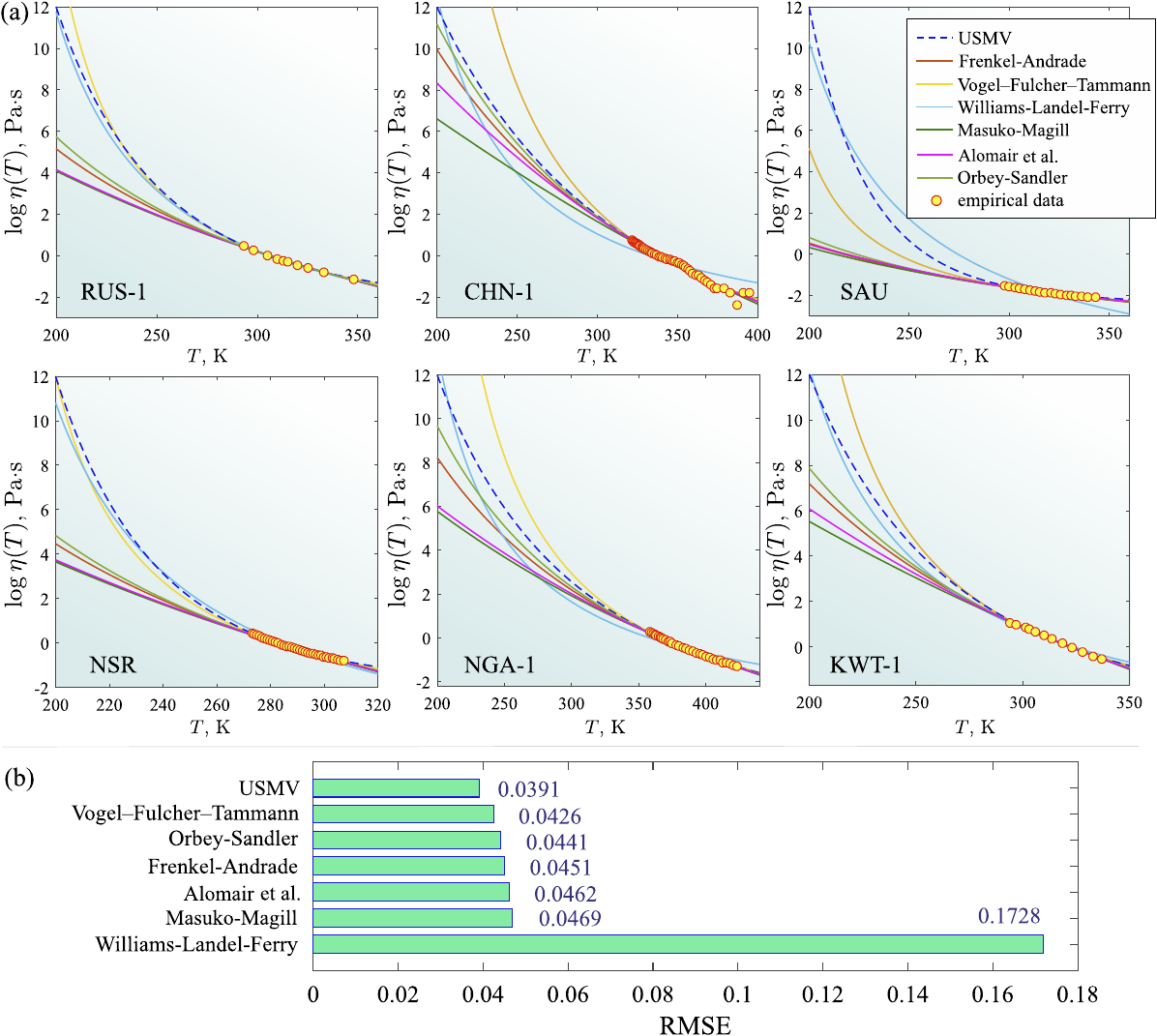}
	\caption{(a) Comparison of the results of different viscosity models applied to empirical crude oil data. Empirical data with the designations RUS-1, CHN-1, SAU, NSR, NGA-1 and KWT-1 are considered. (b) Diagram with the calculated root mean square error (RMSE) between empirical data and viscosity models. The proposed unified scaling model for viscosity (USMV) has the smallest error compared to the other models.}
	\label{fig_5}
\end{figure}

In Figure~\ref{fig_5}, the results of the USMV are compared with other viscosity models. The Vogel-Fulcher-Tamman equation~\cite{Nascimento_Aparicio_2007}, the Frenkel-Andrade relation~\cite{Frenkel_1946,Andrade_1934}, the Masuko-Magill~\cite{Masuko_Magill_1988} and William-Landel-Ferry models~\cite{Kolotova_Kuchina_2018,Williams_Landel_Ferry_1955} were considered, which are the best known of the viscosity models. The Alomair et al.~\cite{Alomair_Jumaa_2016} and Orbey-Sandler~\cite{Bergman_2007} viscosity models are also considered. These models have been proposed to describe the viscosity-temperature data of crude oil and its fractions. All the considered models are compared through fitting the empirical data for the crude oil samples marked as RUS-1, CHN-1, SAU, NSR, NGA-1 and KWT-1 [see Table~\ref{tab_1} and discussion in Supplementary Material]. The description of the empirical data and the estimation of the fitting parameters for these models were performed using the automatic curve fitting function of the MATLAB software (The MathWorks, Inc.,~\cite{Asadi_Matlab_2023}).

Figure~\ref{fig_5}(a) shows that all the viscosity models correctly describe the considered empirical data. Near the liquidus temperature $T_{l}$, these models have similar behavior: practically all the curves converge to a single dependence and correctly extrapolate the empirical data at the high temperatures. These models diverge in the low temperature region (i.e., near the glass transition temperature $T_{g}$). This is mainly due to the fact that some viscosity models are not adapted to describe the viscosity at the temperatures near $T_{g}$. These include the Alomair et al. and Orbey-Sandler viscosity models as well as the Masuko-Magill and Frenkel-Andrade models, which are best suited to describe viscosity near the liquidus temperature. The USMV, the Vogel-Fulcher-Tamman equation and the William-Landel-Ferry model extrapolate the viscosity data most correctly to the viscosity $10^{12}$\,Pa$\cdot$s corresponding to $T_{g}$ [see Figure~\ref{fig_5}(a)]. This is because these models are adapted to reproduce the non-Arrhenius temperature dependence of viscosity. 

Figure~\ref{fig_5}(b) shows that the USMV has the smallest root mean square error (RMSE), $\text{RMSE}\approx0.039$, which is lower compared to the RMSE of other viscosity models for the same empirical data. The RMSE is sensitive to statistical outliers and is generally applied to estimate the prediction error of different models with respect to particular data~\cite{Hyndman_Koehler_2006}. Therefore, the RMSE is suitable for comparing the results of viscosity models with empirical data (see Supplementary Material). This indicates that the USMV generally provides high accuracy compared to other viscosity models over the temperature range covered by the experiment. Such a relatively high accuracy is achieved by ranking the temperature range from $T=0$ to $T=T_{l}$ in a universal way for all oil samples [via the parameters $K_{1}(T_{g},T_{l})$ and $K_{2}(T_{g},T_{l})$] taking into account the glass transition temperature $T_{g}$ and the liquidus temperature $T_{l}$. The largest error is obtained for the Williams-Landel-Ferry model, where $\text{RMSE}\approx0.173$. This model is better suited to reproduce the viscosity at relatively low temperatures (mainly near $T_{g}$)~\cite{Ilyin_Arinina_2016}. For the other viscosity models we have $\text{RMSE}\approx0.045\pm0.002$, which is slightly larger than the error obtained for the USMV. The result of the Vogel-Fulcher-Tamman equation can be different for the low temperature region (near $T_{g}$) and for the high temperature region (near $T_{l}$)~\cite{Gao_Jian_2020}. The Frenkel-Andrade model generally describes viscosity data in the temperature range well above $T_g$, while the Masuko-Magill model fits the temperature range from $T_{g}$ to $T_{l}$~\cite{Masuko_Magill_1988,Ilyin_Arinina_2016}.

The proposed unified scaling viscosity model differs from microscopic and compositional models as well as models based on artificial intelligence, in that it does not need to consider field and oil composition specifics to describe empirical data. Microscopic viscosity models are extremely difficult to describe the oil viscosity because it is not possible to take into account the various interatomic interactions and structural characteristics for a mixture consisting of more than a thousand different chemical compounds~\cite{Sturm_2021}. Compositional models of crude oil viscosity can depend on the oil field and thermodynamic conditions~\cite{Liu_Zhao_2023,Elsharkawy_2003}. The models based on artificial intelligence, such as artificial neural networks, the gradient boosting regression tree, the support vector machine, the stochastic gradient decent, need to be trained on a large set of accurate data~\cite{Li_Zhang_2023,Sun_Huo_2023}. It is necessary to determine a reliable correlation between thermodynamic conditions, composition and viscosity of crude oil~\cite{Aghbashlo_2021}. The USMV is very useful in describing the viscous properties of a system, especially under those conditions (temperature and pressure), where it may be difficult to obtain empirical data for these properties. The ability of the proposed model to predict viscosity over a wide range of temperatures is the result of the temperature scaling concept, which is realized for the temperature range from ultra-low temperatures comparable with $T=0$ to temperatures corresponding with the equilibrium melt. The temperature scaling realizes the transition from absolute temperature to reduced temperature $\widetilde{T}$ and allows one ranked the temperature range $(0; T_l]$ in a uniform manner. The scaling concept in the description of transport characteristics of liquids have significant differences from any microscopic models~\cite{Zaccone_2023,Aitken_Volino_2021} as well as kinetic models~\cite{Avramov_2005,Zheng_Mauro_2011,Mashanov_Darmaev_2022}. Note that the scaling concept does not initially aim to describe viscosity in terms that characterize the physics of viscous flow of a system. Instead, this concept allows one to adequately identify the general regularities in the investigated phenomenon.

\section{Conclusions}

In the present work, it is shown for the first time that the viscosity model realized within the framework of the temperature scaling concept is able to correctly reproduce the temperature dependence of the crude oil viscosity. This model is implemented over a wide temperature range, including temperatures corresponding to the amorphous state of crude oil and temperatures, at which all oil fractions are in the liquid state. The relationship between the parameters of the unified scaling model for viscosity, the glass transition temperature, the fragility index, the activation energy and the API gravity is established for the first time. Thus, it is shown that the parameters of the proposed viscosity model are directly related to the physical characteristics of crude oil, including those that allow the estimation of the glass-forming ability. The accuracy of the proposed model is higher compared to the results of other viscosity models such as the Vogel-Fulcher-Tamman equation, the Frenkel-Andrade equation, the Masuko-Magill and William-Landel-Ferry models as well as the Alomair et al. and Orbey-Sandler viscosity models, which have been developed directly for predicting the viscosity of petroleum products.

\section*{Acknowledgement}
\noindent This work is supported by the Kazan Federal University Strategic Academic Leadership Program (PRIORITY-2030). The authors are grateful to Dmitry Ivanov (Kazan Federal University) for useful discussions. 

\section*{List of notations}

{\footnotesize\tabcolsep1.5pt
	\begin{tabular}{ll}	
		USMV & Universal Scaling Model for Viscosity \\
		API & American Petroleum Institute \\
		RMSE & Root mean square error \\
		$\eta$ & Dynamic viscosity \\
		$T$ & Temperature \\
		$p$ & Pressure \\
		$T_g$ & Glass transition temperature \\
		$T_l$ & Liquidus temperature \\
		$\widetilde{T}$ & Reduced temperature \\
		$\alpha$ & Coefficient in equation of the USMV \\
		$\beta$ & Exponent in equation of the USMV  \\
		$K_{1}$ and $K_{2}$ & Coefficients in scaling temperature concept \\
		$\eta_{\infty}$ & High temperature limit of the viscosity \\
		$E_{A}^{(T_g)}$ & Activation energy at $T_{g}$ \\
		$m$ & Fragility index \\
		$R$ & Universal gas constant \\
		$a_{0}$, $a_{1}$, $a_{2}$, $a_{3}$ & Coefficients of the regression model \\
		$D$ & Self-diffusion coefficient \\
		$\xi$ & Exponent in the generalized Stokes-Einstein relation \\ 
		$C$ & Constant in the generalized Stokes-Einstein relation \\ 
\end{tabular}}

\bibliographystyle{unsrt}

\end{document}